\documentclass[pdflatex,sn-mathphys-num]{sn-jnl}

\usepackage{graphicx}%
\usepackage{multirow}%
\usepackage{amsmath,amssymb,amsfonts}%
\usepackage{amsthm}%
\usepackage{mathrsfs}%
\usepackage[title]{appendix}%
\usepackage{textcomp}%
\usepackage{manyfoot}%
\usepackage{booktabs}%
\usepackage{braket}
\usepackage[utf8]{inputenc}
\DeclareUnicodeCharacter{2212}{\ensuremath{-}}
\usepackage{algorithm}%
\usepackage{algorithmicx}%
\usepackage{algpseudocode}%
\usepackage{listings}%
\usepackage[utf8]{inputenc}
\usepackage[table]{xcolor}
\usepackage{array}
\definecolor{tablemustard}{rgb}{0.9, 0.7, 0.0}
\usepackage{nicematrix, tikz}
\usepackage{arydshln} 
\usepackage{xcolor}   
\usepackage{graphicx}

\usepackage{diagbox}

\usepackage{amsthm}
\usepackage{lmodern}

\begin{document}

\title[Quantum Cryptanalysis of GFSPX]{\centering Quantum Circuit Realization and Grover Cryptanalysis of the Hybrid ARX-SPN Cipher GFSPX}


\author*[1,3]{\fnm{Ibrahim} \sur{Ulgen}}\email{ibrahim.ulgen@siirt.edu.tr}

\author*[2]{\fnm{Hasan Ozgur} \sur{Cildiroglu}}\email{cildiroglu@ankara.edu.tr}

\author[1]{\fnm{Oğuz} \sur{Yayla}}\email{oguz@metu.edu.tr}

\affil[1]{\orgdiv{Institute of Applied Mathematics}, \orgname{Middle East Technical University}, \orgaddress{\street{Universiteler}, \city{Çankaya}, \postcode{06800}, \state{Ankara}, \country{Türkiye}}}

\affil[2]{\orgdiv{Physics Department}, \orgname{Ankara University}, \orgaddress{\street{Dogol st}, \postcode{06100}, \state{Ankara}, \country{Türkiye}}}

\affil[3]{\orgdiv{Department of Mathematics }, \orgname{Siirt University}, \orgaddress{\street{Universite st}, \postcode{56100}, \city{Siirt}, \country{Türkiye}}} 

\abstract{
The security of classical symmetric-key primitives is fundamentally challenged by the emergence of quantum computing, necessitating a rigorous evaluation of their post-quantum resilience. This paper presents a comprehensive quantum circuit realization and Grover cryptanalysis of GFSPX, a lightweight block cipher featuring a 64-bit data block and a 128-bit secret key. GFSPX utilizes a unique hybrid architecture that integrates a 4-branch generalized Feistel structure with both Addition-Rotation-XOR (ARX) and Substitution-Permutation Network (SPN) components. Our quantum implementation optimizes resource distribution by exploiting the inherent reversibility of the Feistel network and employing a compact ripple-carry adder for the ARX layers. The proposed architecture achieves a qubit-optimized footprint of 209 qubits with a baseline quantum cost of 32,498 and a circuit depth of 7,617. To evaluate the cipher’s resistance against quantum adversaries, we construct a parallelized Grover oracle using three plaintext-ciphertext pairs to eliminate spurious matches. Our analysis reveals that the total quantum cost of a key-recovery attack on GFSPX is $1.12 \times 2^{159}$ quantum gates. Although this cost falls below the NIST Level 1 security threshold of $2^{170}$, the hybrid ARX-SPN design demonstrates a higher quantum attack resistance among other lightweight designs. These findings provide critical insights into the balance between classical efficiency and quantum resilience in next-generation cryptographic designs for resource-constrained environments.}

\keywords{GFSPX, Lightweight Cryptography, Quantum Circuit Implementation, Grover's Algorithm, ARX-SPN, Post-Quantum Security}

\maketitle 

\section{Introduction}\label{sec1}

Cryptography provides a formal framework to ensure secure communication, authentication, and data integrity. Conventionally, the security of cryptographic primitives is grounded in the computational intractability of specific mathematical problems. The emergence of quantum information processing challenges classical security models, as the coherent manipulation of quantum states provides physical mechanisms to overcome established limits \cite{arute2019, googleai2025, gidney2025, cain2026, babbush2026, gu2026, webster2026, tripier2026}. Shor’s algorithm \cite{shor1997} compromises widely deployed public-key schemes by efficiently solving integer factorization and discrete logarithms \cite{rivest1978method, diffie1976new, miller1985use, koblitz1987elliptic}. Grover’s algorithm \cite{grover1996} yields a quadratic speedup for unstructured search, effectively halving the security margin of symmetric-key constructions \cite{yamamura2000}. Consequently, the resilience of existing cryptographic primitives necessitates rigorous re-evaluation within frameworks that explicitly exploit quantum resources \cite{daemen2001, li2022, selinger2013quantum}.

In this post-quantum landscape, symmetric block ciphers (BCs) remain fundamentally viable. Their resilience relies on iterative, key-dependent substitutions and permutations that establish rigorous confusion and diffusion \cite{shannon1949communication}. The Feistel network constitutes a core design paradigm for symmetric BCs, enabling block splitting, subkey injection, and iterative structures to perform both encryption and decryption with the same key \cite{Feistel1973, DES1999}. For resource-constrained environments, such as the Internet of Things (IoT) and RFID systems, lightweight BCs are specifically designed to strike a balance between robust security and hardware efficiency \cite{aboushosha2020, khan2025comprehensive, seo2018compact, Cildiroglu_2025a, Cildiroglu_2025b}. Among the core design paradigms, Addition-Rotation-XOR (ARX) and Substitution-Permutation Networks (SPN) are particularly prominent. ARX designs employ simple operations to achieve non-linearity without relying on pre-defined $S$-boxes, reducing storage overhead, while SPN architectures provide rapid diffusion \cite{huang2020automatic, qin2022towards}.

GFSPX is a lightweight cipher that synthesizes a 4-branch Generalized Feistel Structure (GFS) with a hybrid ARX-SPN paradigm \cite{zhang2024gfspx}. GFSPX utilizes two distinct functional blocks: an ARX-based function ($F_1$), designed to achieve non-linearity, and a 32-bit SPN structure ($F_2$), which incorporates the PRESENT's $S$-box and bit-level permutations to facilitate rapid and comprehensive diffusion across all data paths \cite{bogdanov2007}. This hybrid approach enables GFSPX to achieve a full avalanche effect in six rounds, addressing the slow diffusion properties of pure ARX ciphers while maintaining a hardware footprint comparable to standardized lightweight primitives \cite{dobraunig2021ascon}. The inclusion of modular addition in this hybrid design further motivates an investigation into its impact on Grover's attack complexity.

Quantifying the post-quantum resilience of a cryptographic primitive fundamentally requires a detailed mapping into a quantum circuit \cite{nist2016, nist2020}. Such a realization enables the construction of the Grover oracle and facilitates a precise estimation of the quantum resources used to mount a key-recovery attack. In this work, we present a comprehensive quantum circuit realization of GFSPX. Our implementation focuses on minimizing the qubit count and circuit depth by exploiting the reversibility of the Feistel structure and optimizing the gate-level architecture of both the modular addition and substitution layers. We demonstrate a resource-efficient design that utilizes 209 qubits and achieves an optimized gate-to-depth ratio of 4.27, making it suitable for evaluation on current and near-term quantum devices. Our analysis reveals that the modular addition primitive accounts for the majority of the total gate count, despite contributing minimally to the overall circuit depth.

Furthermore, we quantitatively evaluate the quantum security of GFSPX against Grover's algorithm within the NIST \texttt{MAXDEPTH} framework \cite{nist2016}. To eliminate spurious matches in the 128-bit key space, we construct a parallelized Grover oracle utilizing three plaintext-ciphertext pairs. The resulting attack requires 628 qubits and exhibits a total quantum cost of approximately $1.12 \times 2^{159}$ with its comparison with respect to other lightweight ciphers. Although this estimated cost falls below NIST's Level 1 security threshold of $2^{170}$ quantum gates, our analysis indicates that GFSPX does not meet the strict post-quantum criteria for security against a sufficiently advanced quantum adversary. However, its optimized quantum circuit implementation and relatively high temporal depth offer a competitive quantum cost compared to other lightweight BCs, providing valuable insights for future quantum-resistant lightweight designs. Moreover, our resource estimation confirms that the modular addition-based ARX primitive relatively increases the Grover attack complexity. These findings provide insights for the development of future quantum-resistant lightweight cryptographic designs.

The remainder of this paper is organized as follows: Section \ref{sec2} details the classical structure of GFSPX, including its hybrid ARX-SPN round functions and key schedule. Section \ref{sec3} describes the proposed quantum circuit implementation of the cipher. Section \ref{sec4} provides a comprehensive quantum resource estimation and comparison with other lightweight BCs. Finally, Section \ref{sec5} analyzes the security of GFSPX against Grover's algorithm and concludes with resource estimates under the \texttt{MAXDEPTH} constraint.

\section{The structure of GFSPX}\label{sec2}

\begin{figure}[h!]
\centering
\includegraphics[width=\textwidth]{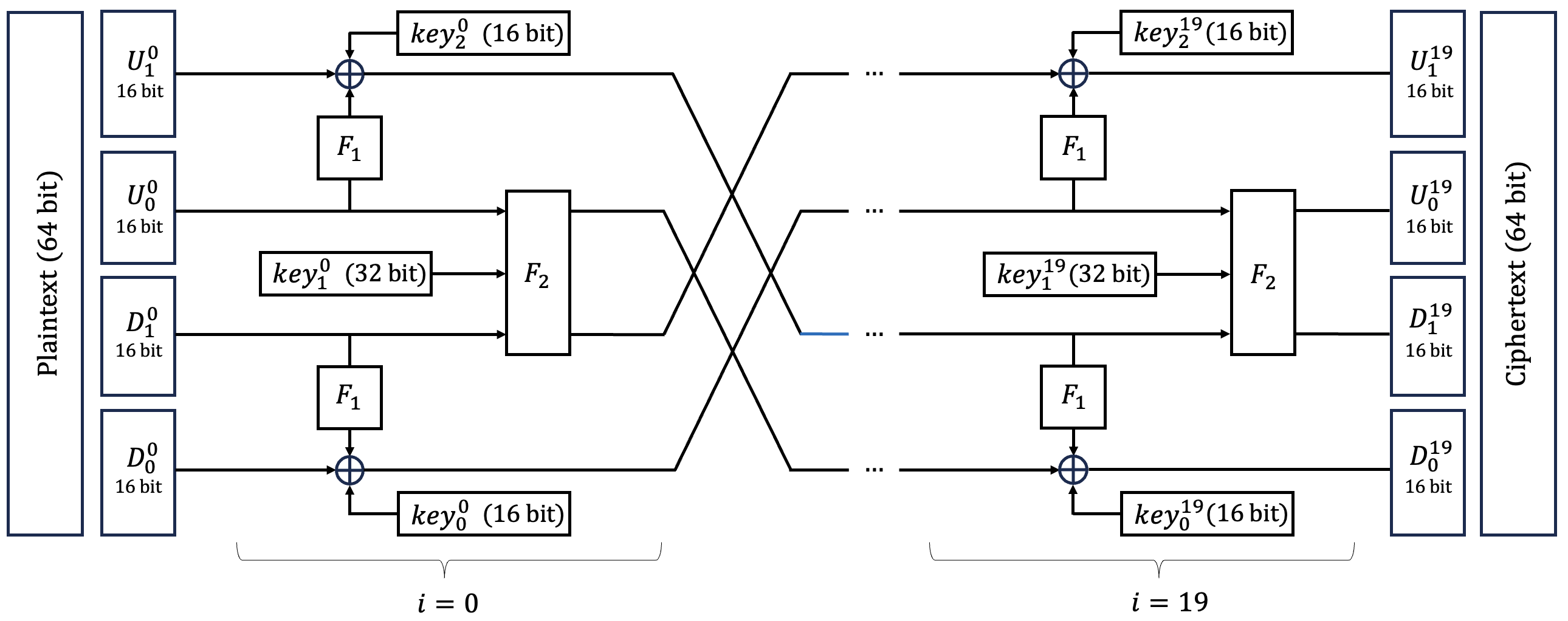}\vspace{5mm}
\includegraphics[width=\textwidth]{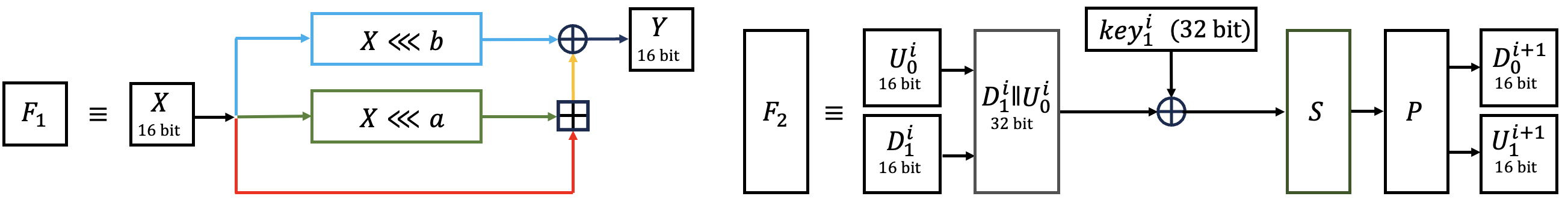}
\caption{\centering Core round functions of GFSPX \cite{zhang2024gfspx}: \\ $D_{1}^{i+1}= F_{1}(U_{0}^{i}) \oplus U_{1}^{i} \oplus key_{2}^{i}$, $U_{0}^{i+1} = F_{1}(D_{1}^{i}) \oplus D_{0}^{i} \oplus key_{0}^{i}$, and $(U_{1}^{i+1}, D_{0}^{i+1})= F_{2}\big( D_{1}^{i} \,\|\, U_{0}^{i}, \, key_{1}^{i}\big)$}
\label{fig3}
\end{figure}

GFSPX is a symmetric-key LBC designed for resource-constrained IoT environments, operating on 64-bit data blocks with a 128-bit secret key \cite{zhang2024gfspx} (See Fig. \ref{fig3}). The encryption process consists of 20 rounds, using a hybrid architecture that integrates a 4-branch generalized Feistel structure with ARX-SPN components. 64-bit plaintext is divided into four 16-bit branches, $\lbrace D_0, D_1, U_0, U_1 \rbrace$, ordered from the most significant bits (MSB) to the least significant bits (LSB). These segments are subsequently processed through two distinct nonlinear functional blocks. To ensure adequate diffusion across all data paths, a branch permutation is applied at the end of each cycle, except for the final round.

\vspace{1mm}

\noindent \textbf{The Round Function:} The GFSPX round function is characterized by the integration of the $F_1$ and $F_2$ functional blocks. The function $F_1: \{0,1\}^{16} \to \{0,1\}^{16}$ is defined as $F_1(X) = \bigl(X \boxplus (X \lll a)\bigr) \oplus (X \lll b), \text{with } a = 5,\ b = 1.$ The evaluation of $F_1$ proceeds in three sequential stages. First, two linear shifts are performed, yielding $R_1 = X \lll 5$ and $R_2 = X \lll 1$. Second, a modular addition is carried out, $S = X \boxplus R_1 = (X + R_1) \bmod 2^{16}$, which constitutes the sole source of non-linearity within $F_1$. Third, the output is obtained as $Y = S \oplus R_2$. Here, $\oplus$ denotes bitwise XOR, $\boxplus$ stands for modular addition over $\mathbb{Z}/2^{n}\mathbb{Z}$, and $X \lll t$ represents the left circular (cyclic) shift of $X$ by $t$ bit positions. Complementing the ARX operations, the $F_2$ function is designed as a 32-bit SPN structure. The process begins with the bitwise XOR of the input and a 32-bit subkey ($key_1^i$), followed by a substitution layer comprising eight parallel 4-bit $S$-boxes consistent with the $S$-box definitions specified in the PRESENT \cite{bogdanov2007}, see Table \ref{$S$-box}. Last, a bit-level permutation layer ($P$-layer) is applied to facilitate effective diffusion throughout the internal state (Table \ref{Permutation}). By integrating this hybrid design, GFSPX achieves a full avalanche effect in just six rounds \cite{zhang2024gfspx}.

\begin{table}[t]
\centering
\caption{$S$-box (Substitution-box) of GFSPX}
\label{$S$-box}
\begin{tabular}{c|cccccccccccccccc}
$x$ & 0 & 1 & 2 & 3 & 4 & 5 & 6 & 7 & 8 & 9 & A & B & C & D & E & F \\ \hline
$S(x)$ & C & 5 & 6 & B & 9 & 0 & A & D & 3 & E & F & 8 & 4 & 7 & 1 & 2
\end{tabular}
\end{table}

\begin{table}[h]
\centering
\caption{Permutation table of GFSPX}
\label{Permutation}
\begin{tabular}{c c @{\qquad\vrule\quad} c c @{\qquad\vrule\quad} c c @{\qquad\vrule\quad} c c}
$i$ & $P(i)$ & $i$ & $P(i)$ & $i$ & $P(i)$ & $i$ & $P(i)$ \\
\hline
0  & 0  & 8  & 2  & 16 & 4  & 24 & 6  \\
1  & 8  & 9  & 10 & 17 & 12 & 25 & 14 \\
2  & 16 & 10 & 18 & 18 & 20 & 26 & 22 \\
3  & 24 & 11 & 26 & 19 & 28 & 27 & 30 \\
4  & 1  & 12 & 3  & 20 & 5  & 28 & 7  \\
5  & 9  & 13 & 11 & 21 & 13 & 29 & 15 \\
6  & 17 & 14 & 19 & 22 & 21 & 30 & 23 \\
7  & 25 & 15 & 27 & 23 & 29 & 31 & 31 \\
\end{tabular}
\end{table}

\begin{figure}[h]
\centering
\includegraphics[width=0.9\textwidth]{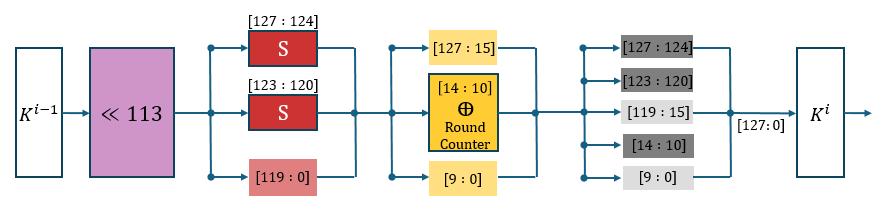}
\caption{\centering Key scheduling mechanism of GFSPX.}
\label{fig2}
\end{figure}

\noindent \textbf{Key Schedule}: The key scheduling algorithm operates on a 128-bit master key register $K^0 = (k_{127},\ldots,k_{0})$ from MSB to LSB (see Fig. \ref{fig2}). The initial round subkeys $key_{0}^{i}$, $key_{1}^{i}$, and $key_{2}^{i}$ are directly extracted as contiguous segments of $K^0$: $key_{0}^{0}=K^0[127:112]$, $key_{1}^{0}=K^0[111:80]$, and $key_{2}^{0}=K^0[79:64]$. Following subkey extraction, the register is transformed for the subsequent iteration through a series of operations. First, the register undergoes a left cyclic shift of 113 bits (${\rm temp}_0 = K^{i-1} \lll 113$ for $i=1,2,\ldots,19$). Next, the 8 MSBs are processed through the $S$-box in two 4-bit nibbles: ${\rm temp}_1 = \mathcal{S}({\rm temp}_0[127:124])$ and ${\rm temp}_2 = \mathcal{S}({\rm temp}_0[123:120])$. Subsequently, a round counter ($RC$) is XORed with a specific bit-slice, targeting ${\rm temp}_0[14:10]$, to produce the updated segment ${\rm temp}_3 = {\rm temp}_0[14:10] \oplus RC$. The succeeding key state, $K^{i}$, is constructed by concatenating these processed segments: $K^{i} = {\rm temp}_1 ~ || ~ {\rm temp}_2 ~ || ~ [119:15] ~ || ~ {\rm temp}_3 ~ || ~ [9:0]$ for $i=1,2,\ldots,19$.

The decryption process relies on the inherent reversibility of the generalized Feistel structure. While the ARX-based function $F_1$ is applied directly, the SPN-based $F_2$ function requires the application of its inverse, $F_2^{-1}$. This inversion incorporates $S^{-1}$-box and $P^{-1}$-layer, which are derived by reversing their respective mappings as defined in Tables \ref{$S$-box} and \ref{Permutation}. Furthermore, the subkeys derived from the key schedule are introduced in reverse order to recover the original plaintext.

\section{Quantum Realization of the GFSPX}\label{sec3}

Evaluating the security of the GFSPX against Grover attacks requires an estimation of the quantum resources. Since the complexity of such attacks is fundamentally governed by the quantum gate cost and circuit depth of the target primitive, constructing an optimized quantum implementation is a prerequisite for rigorous security analysis. Therefore, we present the quantum realization of GFSPX by mapping its classical structure into a reversible quantum circuit. To minimize key quantum resources, we exploit the inherent reversibility of the Feistel network and optimize the gate-level architecture of the ARX and SPN components.

\begin{figure}[h!]
    \centering
    \includegraphics[width=\textwidth]{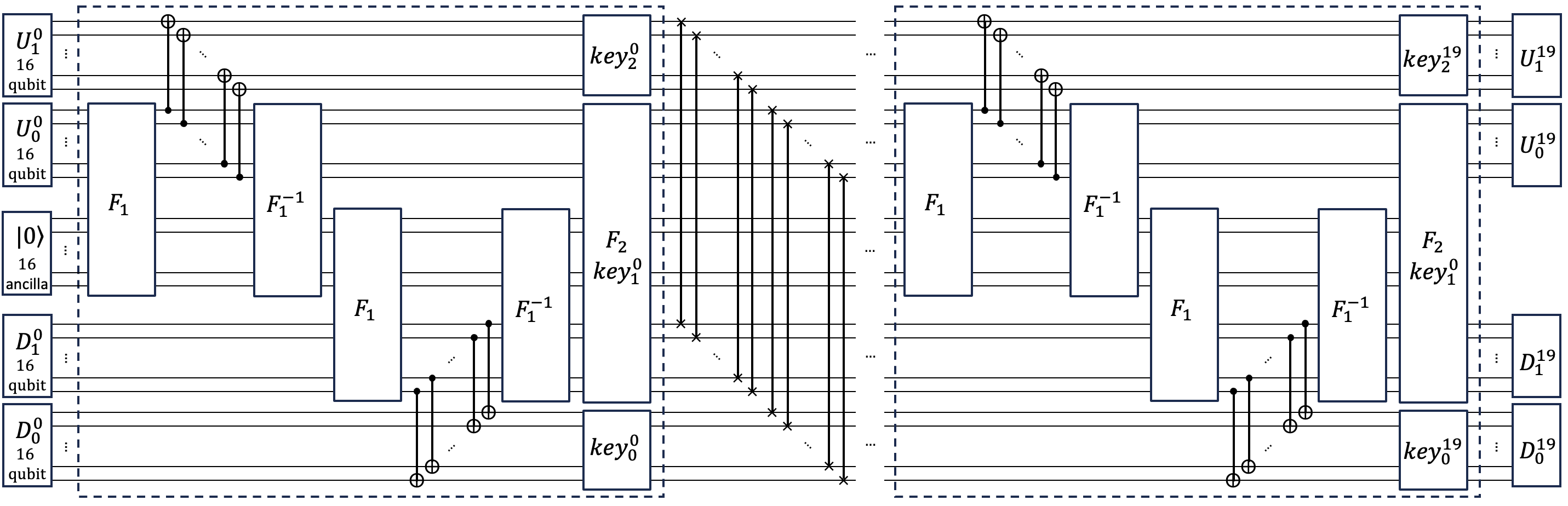}
    \caption{\centering The complete quantum circuit for the GFSPX}
    \label{fig:placeholder}
\end{figure}

The proposed quantum architecture (Fig. \ref{fig:placeholder}) operates on a 64-qubit input, which is initially divided into four 16-qubit subsystems, $\lbrace D^{0}_0, D^{0}_1, U^{0}_0, U^{0}_1 \rbrace$, ordered from the most significant qubits (MSQ) to the least significant qubits (LSQ). It incorporates an auxiliary 16-qubit register to store intermediate states during the $F_1$ operations. Within a given round $i$, the $F_1$ operator first acts on the $U^{i}_0$ subsystem using the ancilla qubits, and the outcome is CNOTed into the $U^{i}_1$ register. To recover the initial $U^{i}_0$ and ancillaries $|0\rangle^{\otimes 16}$, a subsequent uncomputation $F^{-1}_1$ is performed. This sequence is then applied to the $D^{i}_1$ subsystem. Following these operations, the $F_2$ operator is applied to both $U^{i}_0$ and $D^{i}_1$ subsystems, while the round-specific subkeys derived from the key schedule ($key^{i}_2, key^{i}_1, \text{and } key^{i}_0$) are simultaneously CNOTed into the $U^{i}_1$, $F_2$, and $D^{i}_0$ layers. Thus, the $i$-th round is completed. To ensure diffusion before the subsequent round commences, a permutation of the subsystems is executed, where the $U^{i}_1$ and $U^{i}_0$ registers are mapped to $D^{i+1}_1$ and $D^{i+1}_0$, respectively. This protocol is applied in all rounds except for the final round, which produces the ciphertext.

\begin{figure}[h!]
\centering
\includegraphics[width=\textwidth]{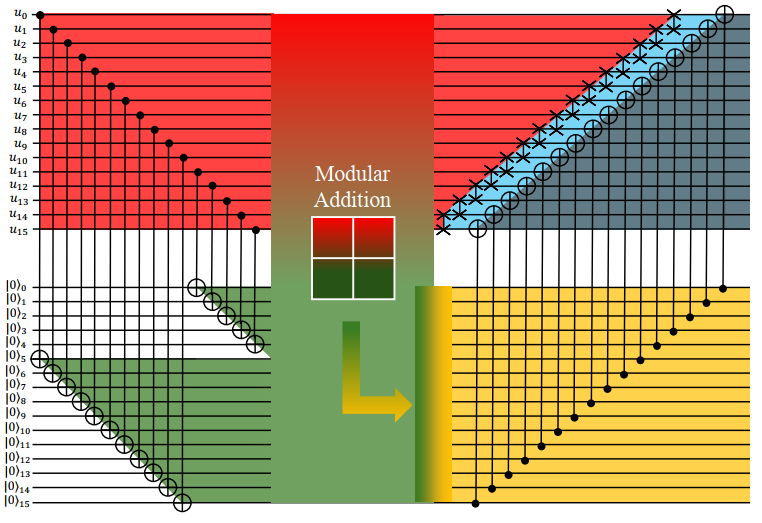}
\caption{\centering Quantum realization of the ARX-based $F_1$ function.}
\label{fig4}
\end{figure}

\textbf{$F_1$ function (16-qubit ARX)}: The quantum circuit for the $F_1$ function is designed to implement $F_1(X)=(X\boxplus(X \lll 5))\oplus(X\lll 1)$ while prioritizing resource efficiency. As illustrated in Fig. \ref{fig4}, the process initiates with the preparation of operands for 16-bit modular addition. Specifically, the primary register $X$ serves as the control to map the 5-bit left-cyclically shifted state, $X \lll 5$, onto a 16-qubit ancilla target initialized to $|0\rangle^{\otimes 16}$ via bitwise CNOT operations. 

\textit{The n-bit modular addition (The quantum RCA)}: The overall computational complexity of the $F_1$ operator is primarily driven by its arithmetic layer, we employ the quantum ripple-carry adder (RCA) architecture proposed by Cuccaro, Draper, Kutin, and Moulton \cite{cuccaro2004}. This approach offers a linear-depth solution that minimizes qubit overhead, making its minimal-ancilla property particularly well-suited for our proposed architecture. 

The efficiency of the quantum RCA stems from a symmetric, two-phase computational structure facilitated by two compact subroutines. The first is the in-place majority (MAJ) gate, which computes $c_{i+1} = \mathrm{MAJ}(a_i, b_i, c_i) = a_i b_i \oplus a_i c_i \oplus b_i c_i$ using two CNOT gates followed by a single Toffoli gate, where $a_i$ and $b_i$ denote the $i$-th bits of the n-bit input operands $a$ and $b$, and $c_i$ represents the carry bit propagated from the preceding bit position; this construction advances the carry chain in the forward pass without consuming additional ancillae. The second is the UnMajority-and-Add (UMA) gate, which performs three operations within a single compact structure: it reverses the action of the MAJ gate to restore the input bit $a_i$ to its original location, returns the carry $c_i$ to the preceding register, and simultaneously writes the sum bit $s_i = a_i \oplus b_i \oplus c_i$ to the corresponding output position. The forward pass thus propagates the carry chain from the least significant to the most significant bit, while the backward pass uncomputes the intermediate carries and produces the sum bits in a reverse-order ripple, ensuring that the entire computation remains reversible.

In its standard form, the Cuccaro RCA achieves a depth of $2n + 4$ using $2n - 1$ Toffoli and $5n - 3$ CNOT gates. For the present application, we adopt the modulo-$2^n$ variant, where the most significant carry bit $c_n$ is omitted as it is irrelevant for modular addition. This modification reduces the resource requirements to $2n - 3$ CCNOT gates, $5n - 7$ CNOT gates, and $2n-6$ NOT gates, with a reduced depth of $2n + 2$. 

In our 16-qubit implementation, this design choice yields a concrete resource budget of 29 CCNOT gates, 73 CNOT gates, 26 NOT gates, and a circuit depth of 34 for the arithmetic layer. Beyond reducing the qubit count, omitting the terminal carry bit eliminates the final Toffoli operation, directly enhancing the temporal efficiency of the $F_1$ operator. The remaining linear components—circular shifts and bitwise XORs—are integrated via SWAP and CNOT gates, respectively (see Fig. \ref{fig4}). Accordingly, the total gate complexity for the $F_1$ operator is characterized by 26 NOT, 105 CNOT, 29 CCNOT, and 15 SWAP gates.

\textbf{$F^{-1}_1$ function (16-qubit ARX)}: The inverse of $F_1$ uncomputes the function to restore the initial state, thereby inherently preserving the gate configuration and architectural complexity of the forward operation. This uncomputation process requires 26 NOT, 105 CNOT, 29 CCNOT, and 15 SWAP gates, mirroring the resource requirements of the $F_1$ operator as detailed in \cite{cuccaro2004}.

\textbf{$F_2$ function (32-bit SPN)}: The $F_2$ function executes a sequential SPN transformation on the recovered $D_1^i$ and $U_0^i$ subsystems. The $F_2$ function executes a sequential SPN transformation on the concatenated 32-qubit state ($D_1^i || U_0^i$). This transformation is realized through three sequential quantum layers:

\textit{Round Subkey Addition}: The 32-qubit round subkey ($key^{i}_1$) is integrated into the 32-qubit concatenated ($D^{i}_1 \parallel U^{i}_0$) via 32 parallel CNOT gates in Fig. \ref{fig7}.

\begin{figure}[H]
\centering
\includegraphics[width=0.48\textwidth]{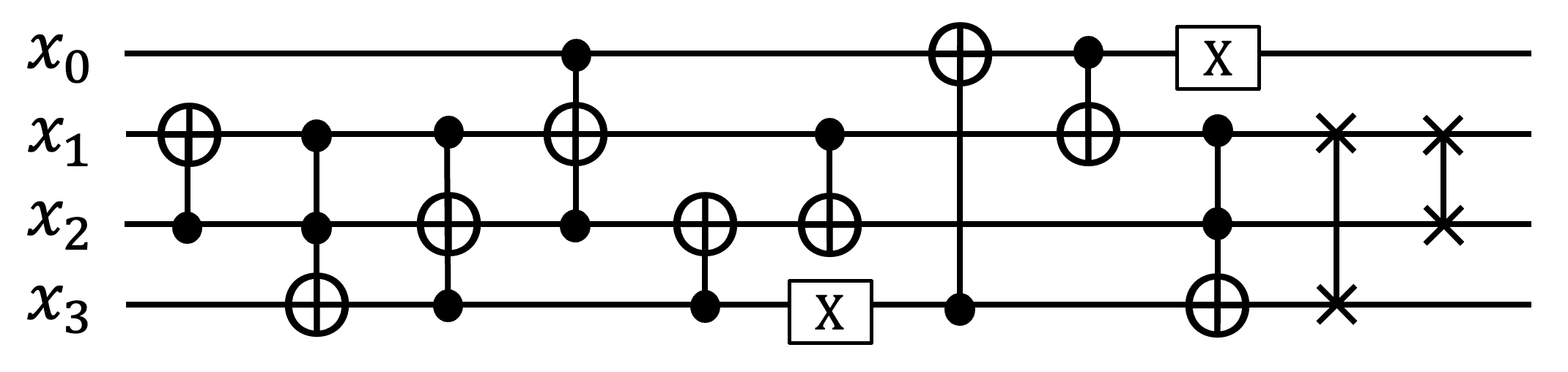}
\includegraphics[width=0.48\textwidth]{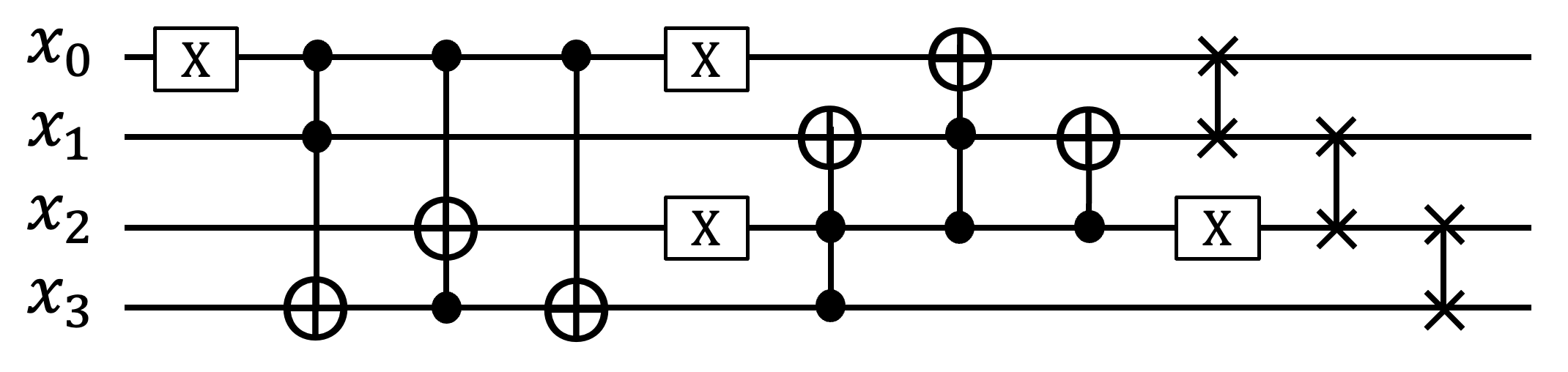}
\caption{\centering Quantum implementation of $S$-box (left) and $S^{-1}$-box (right)} 
\label{fig5}
\end{figure}

\textit{$S$-box}: The $S$-box layer used in GFSPX in Table \ref{$S$-box}, is adopted from the PRESENT algorithm \cite{bogdanov2007} and its quantum implementation has been synthesized by the LIGHTER-R framework \cite{dasu2019, chun2023dorcis, Ozcan2025}. As illustrated in Fig. \ref{fig5}, the quantum implementation of the $S$-box features a specific optimization that decomposes the non-linear permutation into a minimum set of CNOT and Toffoli gates without requiring additional ancilla qubits. The inverse transformation of the $S$-box is illustrated in Fig. \ref{fig5}. The quantum resource requirements for the $S$-box implementation comprise 2 NOT gates, 5 CNOT gates, 4 CCNOT (Toffoli) gates, and 2 SWAP gates. Conversely, the inverse $S$-box ($S^{-1}$) is realized through a configuration of 4 NOT, 2 CNOT, 4 CCNOT, and 3 SWAP gates.

\begin{figure}[H]
    \centering
    \begin{minipage}{0.45\textwidth}
        \centering
        \includegraphics[width=\textwidth]{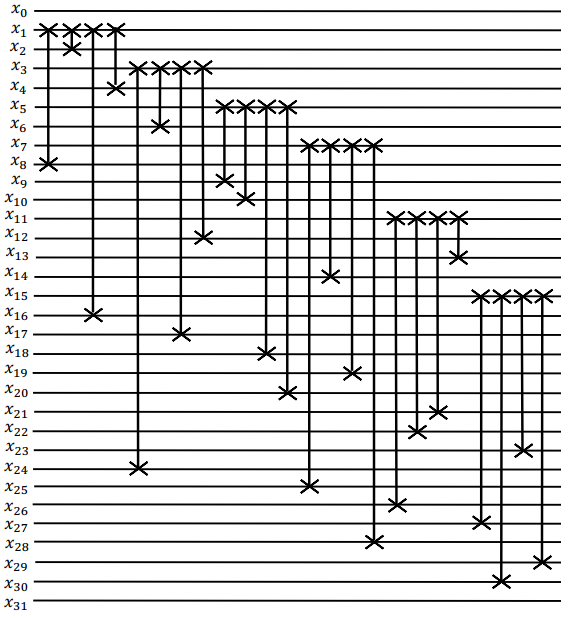}
    \end{minipage}
    \begin{minipage}{0.45\textwidth}
        \centering
        \includegraphics[width=\textwidth]{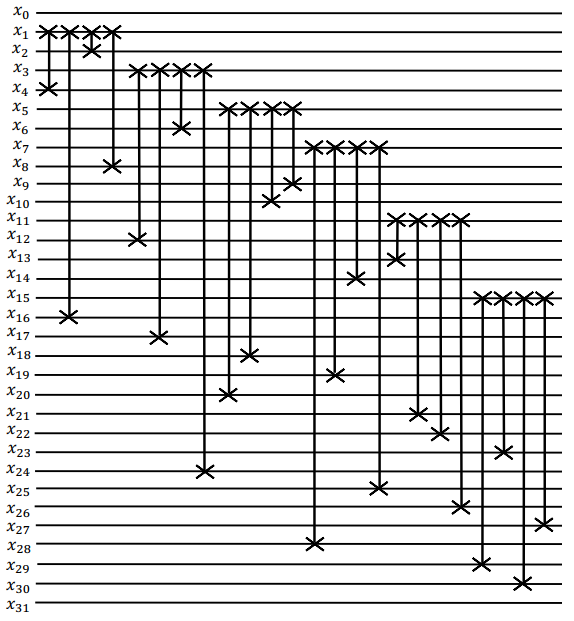}
    \end{minipage}
    \caption{\centering Quantum implementations of (a) $P$-box and (b) $P^{-1}$-box.}
    \label{fig6}
\end{figure}

\textit{$P$-layer}: Both $P$ and $P^{-1}$ layers, detailed in Table \ref{Permutation}, are implemented by systematically reordering the 32-qubit state using SWAP gates (Fig. \ref{fig6}). This 32-bit layer is realized using 24 SWAP gates organized into six parallelizable quartets. Since these quartets are logically independent, they facilitate simultaneous execution across the register. This parallelization strategy is critical for resource optimization as it allows the entire permutation to be executed without increasing the overall circuit depth.

\begin{figure}[H]
\centering
\includegraphics[width=0.8\textwidth]{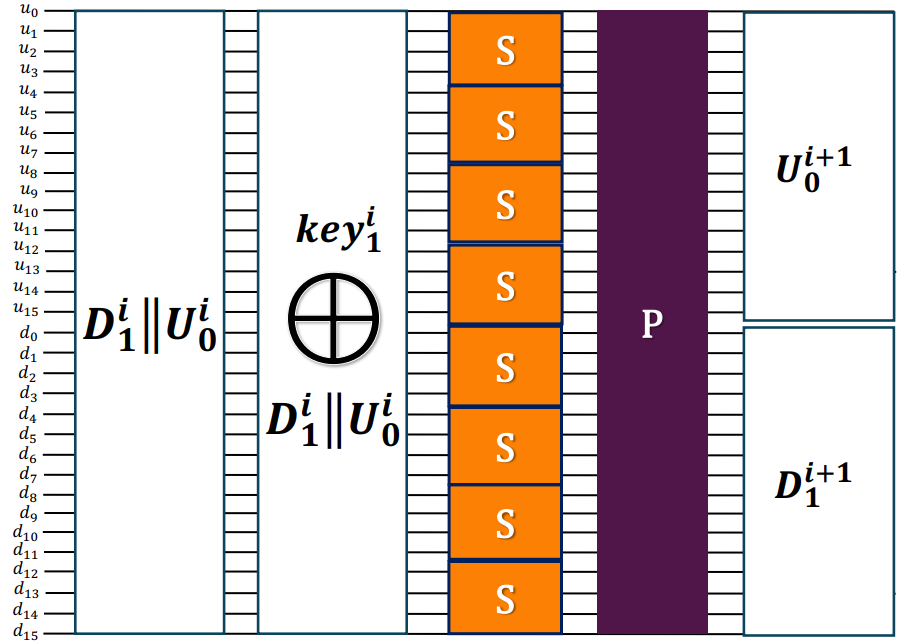}
\caption{\centering Quantum realization of the SPN-based $F_2$ function. It yields a total resource cost of 16 NOT, 72 CNOT, 32 CCNOT, and 40 SWAP gates.}
\label{fig7}
\end{figure}

\begin{figure}[h!]
\centering
\includegraphics[width=\textwidth]{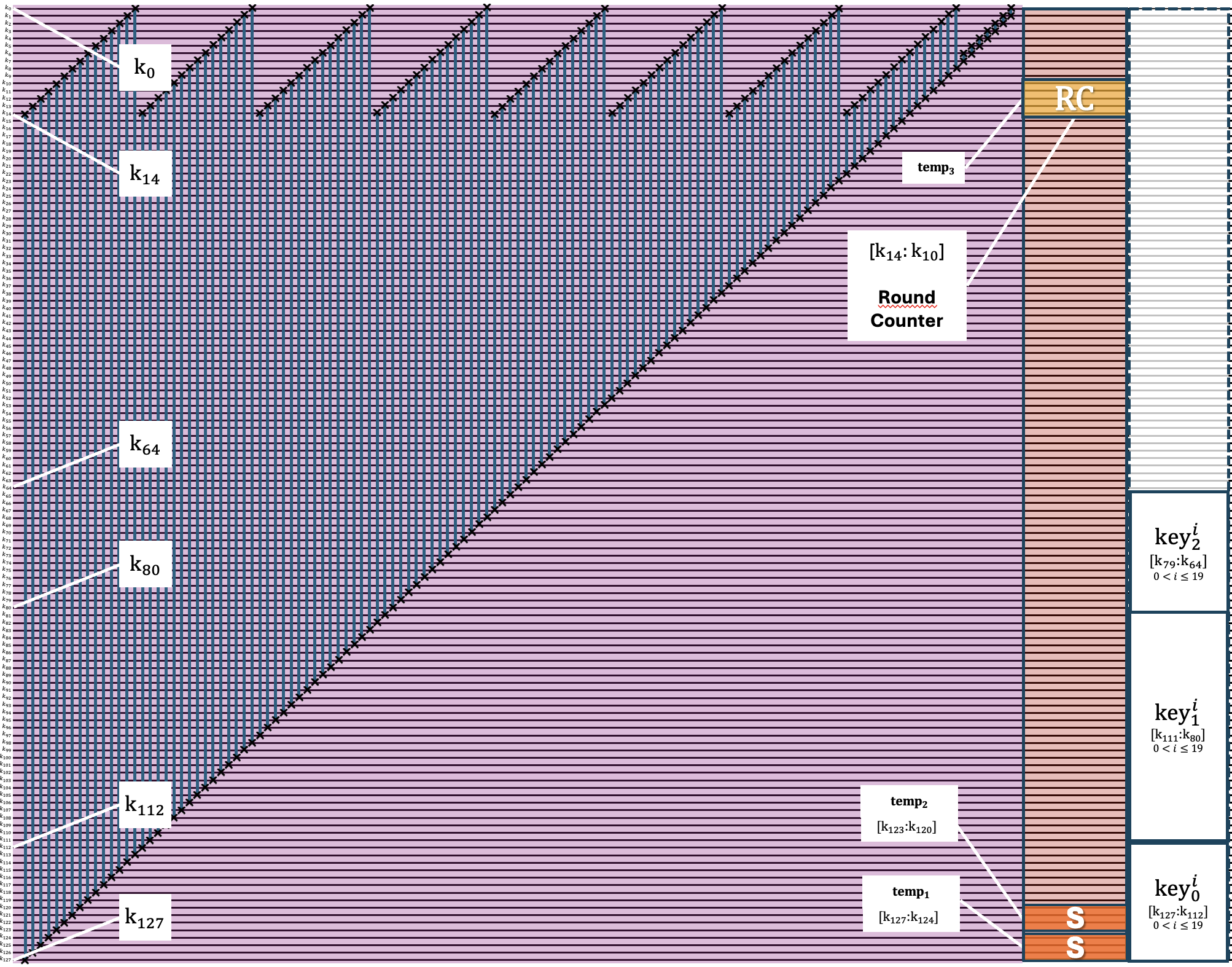}
\caption{\centering Quantum implementation of the GFSPX key schedule.}
\label{fig8}
\end{figure}

\textbf{Dynamic Key-Schedule:} The GFSPX key schedule derives round-specific subkeys by unitarily evolving a 128-qubit master register. It initiates with a 113-bit left circular rotation (LCR) of the 128-qubit register. To minimize circuit depth, this cyclic shift is synthesized using 127 SWAP gates arranged into nine parallel execution layers (see Fig. \ref{fig8}). Following the cyclic shift, the eight MSQ are partitioned into two 4-bit segments ($temp_1, temp_2$) and routed through parallel $S$-boxes to induce non-linearity. Concurrently, a round-counter layer of NOT gates evaluates the $[k_{14}:k_{10}]$ subsystem (Table \ref{round}). The resulting 128-qubit configuration inherently provides the required 64-qubit subkeys, ${key^{i}_0, key^{i}_1, key^{i}_2}$, for rounds $i \in \{1,2, \dots, 19\}$ without additional gate overhead (Fig. \ref{fig8}). The inherent reversibility of the Feistel structure facilitates Grover's oracle, permitting the uncomputation of the schedule to recover the initial key state without auxiliary qubit overhead. Thus, the subkeys are provided in reverse for the decryption process. 

\begin{table}[h!]
\centering
\definecolor{gold}{RGB}{218, 165, 32} 

\resizebox{\textwidth}{!}{
\begin{tabular}{|c|c|c|c|c|c|c|c|c|c|c|c|c|c|c|c|c|c|c|c|c|}
\hline
\rowcolor{gold}
{$k_Q$ ~ \textbackslash ~ RC} & 0 & 1 & 2 & 3 & 4 & 5 & 6 & 7 & 8 & 9 & 10 & 11 & 12 & 13 & 14 & 15 & 16 & 17 & 18 & 19 \\ \hline
\cellcolor{gold}$k_{14}$ & & & & & & & & & & & & & & & & & X & X & X & X \\ \hline
\cellcolor{gold}$k_{13}$ & & & & & & & & & X & X & X & X & X & X & X & X & & & & \\ \hline
\cellcolor{gold}$k_{12}$ & & & & & X & X & X & X & & & & & X & X & X & X & & & & \\ \hline
\cellcolor{gold}$k_{11}$ & & & X & X & & & X & X & & & X & X & & & X & X & & & X & X \\ \hline
\cellcolor{gold}$k_{10}$ & & X & & X & & X & & X & & X & & X & & X & & X & & X & & X \\ \hline
\end{tabular}
}
\caption{\centering Round counter algorithm for GFSPX}
\label{round}
\end{table}

\textbf{The Round Function:} As highlighted by the dashed boundaries in Fig. \ref{fig:placeholder}, each round function is a composition of two $F_1$ operators, their corresponding uncomputation counterparts ($2 \times F_1^{-1}$), a single $F_2$ primitive, and 32 CNOT gates. These components are integrated within a round-dependent sub-key layer that utilizes a dual-path injection strategy. While $key^{i}_0$ and $key^{i}_2$ are applied to the state via 32 CNOT gates, and $key^{i}_1$ is embedded within the $F_2$ functional block, all subkeys are integrated into the circuit simultaneously. To facilitate the iterative execution, each round function, except for the final iteration, is linked with the succeeding round through 32 SWAP gates, as indicated in the circuit diagram, see Fig. \ref{fig:placeholder}.


\begin{table}[h]
\setlength{\tabcolsep}{4pt}
\centering
\caption{Quantum resource requirements for implementation of GFSPX}
\label{table:resource}
\begin{tabular}{@{\hskip 0pt}l@{\hskip 0pt}c@{\hskip 2pt}c@{\hskip 2pt}c@{\hskip 2pt}c@{\hskip 2pt}c@{\hskip 2pt}c@{\hskip 2pt}c@{\hskip 0pt}}
\hline
LAYER & NOT & CNOT & CCNOT & SWAP & TOTAL & \begin{tabular}[c]{@{}c@{}}COST\\ ($\text{COST}^{\spadesuit}$)\end{tabular} & DEPTH \\
\hline
$S$              & 2 & 5 & 4 & 2 & 13  & 31 (37) & 33 \\
$S^{-1}$         & 4 & 2 & 4 & 3 & 13  & 30 (39) & 32 \\
$P~ \& ~ P^{-1}$  & 0 & 0 & 0 & 24 & 24 & 0 (72) & 12 \\
$^{*}K^i$        & 4+RC & 10  & 8  & 131 & 153+RC & 62+RC & 36 \\
    &  &   &  &  &  &  (455+RC) &  \\
$\boxplus ~ \& ~ \boxplus^{-1}$ & 26 & 73 & 29 & 0 & 128  & 273 (273) & 34 \\
$F_{1}~ \& ~ F^{-1}_{1}$ & 26 & 105 & 29 & 15 & 175  & 305 (350) & 81 \\
$F_{2}$          & 16  & 72  & 32  & 40  & 160  & 280 (400) & 52 \\
$F^{-1}_{2}$     & 32 & 48 & 32 & 48  & 160  & 272 (416) & 54 \\ 
\hdashline
Round & 120 & 556 & 148 & 100 & 924 & 1564 & 378 \\
Function & & & & &  & (1864) &  \\
($i=0$) &  &  &  &  &  &  &  \\
\hdashline
Round  & 124+RC & 566 & 156 & 231 & 1077+RC & 1626+RC & \\
Function  &  &  &  &  &  & (2319+RC) & 378 \\
($1 \le i \le 19$)  &  &  &  &  &  &  &  \\
\hline \\
\textbf{GFSPX} & 2516 & 11310 & 3112 & 5097 & 22035 & 32498 & 7617\\
 &  &  &  &  &  & (47789) &  \\
\hline
\end{tabular}
\smallskip
\footnotesize * This represents the computational cost of the 128-bit key schedule for all iterations except the first round. Following the key is generated, subkeys ($key^i_0$, $key^i_1$, $key^i_2$) are systematically sent to the relevant functional units within the round transformation. \\
* The gate complexity of RC is round-dependent, varying throughout the execution. The total complexity for the RC component is established as 40 NOT gates over all iterations. \\
$\spadesuit$ Values enclosed in parentheses represent the circuit analysis scenarios where SWAP gates are explicitly included. To incorporate SWAP gates into the overall cost, each gate is decomposed into three CNOT operations.
\end{table}

\section{Quantum resources for the GFSPX}\label{sec4}

To evaluate the resistance of GFSPX against Grover attacks, we quantify the quantum resource overhead, specifically gate costs and circuit depth, of its oracle. This evaluation decomposes the architecture into its fundamental operators: the $S$-box ($S, S^{-1}$), permutation layers ($P, P^{-1}$), round-key additions ($K^{i}$), and modular arithmetic ($\boxplus, \boxplus^{-1}$). Because the initial round differs in its key integration and swap-layer configuration, its resource cost is evaluated independently from the subsequent iterations. Aggregating these modular costs yields the cumulative quantum footprint for the full 20-round execution.

We calculate the quantum cost using the standard decomposition metric \cite{shende2009cnot}, assigning a unit cost to NOT and CNOT gates, and the CNOT cost of 6 to the Toffoli gate to account for its ancilla-free CNOT equivalent. Because SWAP implementation costs strictly depend on the target hardware's connectivity topology, we evaluate the overall complexity under two bounding paradigms (Table \ref{table:resource}). Excluding routing overhead, the fundamental gate sequence (2,516 NOT, 11,310 CNOT, and 3,112 CCNOT) yields a baseline cost of 32,498. Conversely, strictly penalizing spatial routing by decomposing the 5,097 SWAP gates into three CNOTs each \cite{Nielsen2010} establishes an upper bound of 47,789.

The quantum circuit depth is calculated by compressing independent parallel gates into single temporal steps, assigning a physical depth of seven and an overhead of four ancilla qubits to each Toffoli gate to account for its fault-tolerant Clifford+$T$ synthesis \cite{selinger2013quantum}. Evaluating the fundamental primitives (Table \ref{table:resource}), the $S$-box ($S, S^{-1}$) and $P$-layers span depths of 33, 32, and 12, respectively. Similarly, the round-key addition ($K^{i}$) and ($\boxplus, \boxplus^{-1}$) operations necessitate 36 and 34 layers. Upon aggregation of these component layers, the composite $F_1$ and $F^{-1}_1$ transformations yield a total depth of 81, while the $F_2$ and $F^{-1}_2$ operators establish depths of 52 and 54.

Consequently, the modular contributions mentioned in the previous paragraph establish an execution depth of 378 for a single-round function. Scaling this architecture across the complete 20-round encryption process yields a global circuit depth of 7,617. To contextualize these metrics, Table \ref{tab2} benchmarks the GFSPX implementation against other lightweight BCs based on fundamental constraints. This comparative evaluation highlights that the GFSPX architecture achieves high depth number while offering a competitive quantum cost. Moreover, we see that GFSPX achivies the lowest quantum cost with respect to depth (4.27).


\begin{table}[t]
\centering
\caption{Comparison of quantum implementations with other lightweight BCs}
\label{tab2}
\setlength{\tabcolsep}{1pt}
\begin{tabular*}{\textwidth}{@{\extracolsep{\fill}}l cccccccc l}
\toprule%
CIPHER & QUBIT & NOT & CNOT & CCNOT & TOTAL & DEPTH(D) & COST(C) & C/D & REF \\
\midrule
PIPO-64/128 & 192 & 1,477 & 2,248 & 1,248 & 4,973 & 248 & 11,213\textsuperscript{$ \dagger$} & 45.21 & \cite{Jang2025} \\
DEFAULT-128/128 & 256 & 1,131 & 13,824 & 8,192 & 23,147 & 644 & 64,107\textsuperscript{$ \dagger$} & 99.55 & \cite{Jang2025} \\
RECTANGLE-64/128 & 192 & 667 & 6,264 & 2,400 & 9,331 & 226 & 18,931 & 83.77 & \cite{saravanan2021} \\
GIFT-64/128  & 192 & 3,261 & 1,792 & 1,792 & 6,845 & 308 & 14,013 & 45.50 & \cite{jang2020_gift}  \\
GIFT-128/128  & 256 & 10,953 & 6,144 & 6,144 & 23,241 & 528 & 47,817 & 90.56 & \cite{jang2020_gift}  \\
PRESENT-64/128 & 192 & 1,164 & 4,838 & 2,232 & 8,234 & 311 & 17,162 & 55.18 & \cite{paramasivam2023} \\
PUFFIN-64/128 & 192 & 620 & 3,136 & 3,584 & 7,340 & 353 & 21,676 & 61.41 & \cite{paramasivam2023} \\
LiCi-64/128 & 192 & 379 & 12,900 & 1,232 & 14,511 & 1,210 & 20,671\textsuperscript{$\dagger$} & 17.08 & \cite{jing2023} \\
CHAM-64/128 & 204 & 2,320 & 13,200 & 2,320 & 17,840 & 2,615 & 29,440\textsuperscript{$\dagger$} & 11.26 & \cite{Jang2022} \\
CHAM-128/128 & 292 & 4,880 & 28,760 & 4,880 & 38,520 & 5,307 & 62,920\textsuperscript{$\dagger$} & 11.86 & \cite{Jang2022} \\
HIGHT-64/128 & 228 & 4,496 & 22,614 & 5,824 & 32,934 & 2,479 & 62,054\textsuperscript{$\dagger$} & 25.03 & \cite{Jang2022} \\ 
LEA-128/128 & 388 & 11,152 & 32,616 & 10,248 & 54,016 & 6,505 & 105,256\textsuperscript{$\dagger$} & 16.18 & \cite{Jang2022} \\
SIMON-64/128 & 192 & 1,216 & 7,396 & 1,408 & 10,020 & 2,643 & 17,060\textsuperscript{$\dagger$} & 6.45 & \cite{anand2020} \\
SIMON-128/128 & 256 & 4,224 & 17,152 & 4,352 & 25,728 & 8,427 & 47,488\textsuperscript{$\dagger$} & 5.64 & \cite{anand2020} \\
SPECK-64/128 & 194 & 3,131 & 10,669 & 3,233 & 12,581 & 1,863 & 33,228\textsuperscript{$\dagger$} & 17.84 & \cite{Jang_Speck} \\
SPECK-128/128 & 256 & 7,761 & 25,799 & 7,875 & 41,435 & 4,256 & 80,810\textsuperscript{$\dagger$} & 18.99 & \cite{Jang_Speck} \\
\hline
\textbf{GFSPX-64/128} & 209 & 2,516 & 11,310 & 3,112  & 22,035 & 7,617 & 32,498 & 4.27 & This \\
 &   &   &   &   &  & & ~47,789* & 6.27* &  work \\
\botrule
\end{tabular*}
\footnotesize *SWAP gates are included in the cost using a 3-CNOT decomposition.\\
\footnotesize $\dagger$ {Estimated from referenced gate counts, assigning a cost of 1 to NOT/CNOT and 6 to ancilla-free CCNOT decompositions excluded the SWAP overhead of 3-CNOT.}
\end{table}

\section{Grover Analysis of GFSPX}
\label{sec5}

The quantum resilience of GFSPX is evaluated against Grover's algorithm \cite{grover1996}, which provides a quadratic speedup by reducing the exhaustive key search complexity from $\mathcal{O}(2^k)$ to $\mathcal{O}(2^{k/2})$. The physical feasibility of achieving this speedup depends on the hardware synthesis of the quantum oracle, $U_f$. Operating on a uniform superposition of the 128-bit key space, $U_f$ embeds the reversible GFSPX circuit to evaluate candidate keys against known plaintext-ciphertext pairs, executing a phase inversion exclusively on the target state. Given that subsequent amplitude amplification via the standard diffusion operator ($D$) incurs negligible overhead, the total quantum cost of the Grover iteration ($G = D U_f$) is fundamentally dictated by the architectural complexity of the $U_f$ realization.

To eliminate false positives during key recovery, multiple plaintext-ciphertext pairs ($r$) are required to ensure that the phase inversion is triggered exclusively upon a simultaneous match across all target ciphertexts. Following the strategy defined in \cite{langenberg2020}, this parameter is bounded by the key-to-block size ratio ($r > k/n$). Consequently, for the 128-bit key and 64-bit block of GFSPX, $r=3$ pairs are necessitated. To adhere to NIST's \texttt{MAXDEPTH} constraints \cite{nist2016}, we adopt an inner parallelization strategy incorporating six distinct cipher instances (three GFSPX and three GFSPX$^{-1}$) to ensure that data channels are cleared via uncomputation. 

\begin{figure}
\centering
\includegraphics[width=\textwidth]{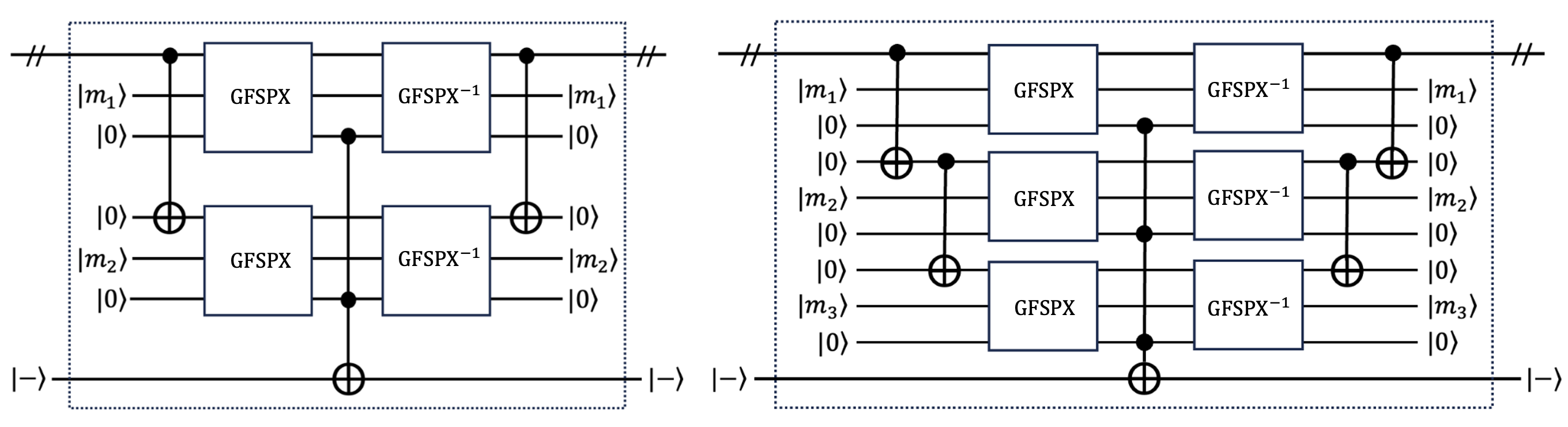}
\caption{\centering  Quantum circuits of Grover oracles $U_f$ for GFSPX with inner parallelization: (left) r=2, and (right) r=3 plaintext-ciphertext pairs} 
\label{Fig13}
\end{figure}

Scaling the base 209-qubit footprint, the composite $U_f$ structure demands $r \cdot 209 + 1 = 628$ qubits (See Fig. \ref{Fig13}). Furthermore, integrating these parallel branches introduces a spatial routing overhead of 704 supplementary CNOT gates, partitioned into $2(r-1)k = 512$ gates to distribute the 128-bit master key and $n \cdot r = 192$ gates to evaluate the 64-bit ciphertext comparisons. The cumulative resource requirements for this integrated oracle are detailed in Table \ref{Grover_resource}.

\begin{table}[h]
\centering
\footnotesize 
\setlength{\tabcolsep}{3pt} 
\caption{Main resource counts for Grover's oracle in the GFSPX}
\begin{tabular*}{\textwidth}{@{\extracolsep{\fill}} l c c c c c c c c @{}}
\toprule
CIPHER & $r$ & QUBIT & NOT & CNOT  & CCNOT & TOTAL & COST & DEPTH \\
       &     &       &     &  &       &  & (COST*) &       \\
\midrule
\textbf{GFSPX} & 2 & 419 & 10,704 & 45,704 & 12,448 & 89,261 & 131,096 & 15,277 \\
      &   &     &       &  &      &  & (192,308) &  \\
\textbf{GFSPX} & 3 & 628 & 16,056 & 68,684 & 18,672 & 123,817 & 196,772 & 15,279 \\
      &   &     &       &  &  &  & (257,984) & \\
\bottomrule
\end{tabular*}
\label{Grover_resource}
\footnotesize *SWAP gates are accounted for in the total cost via a 3-CNOT decomposition.
\end{table}

\begin{table}[h]
\centering
\caption{Resource estimation for Grover's Oracle on GFSPX}\label{oracle}
\begin{tabular}{lcccccccccc}
\toprule
CIPHER [ARCH.]  & $r$ & GATE COST & $\#$GATES & FULL DEPTH & TOTAL COST\\
\midrule
PIPO-64/128 [$\clubsuit$] & 2 & ${*}$ 45,364  & ${*}$1.33 $\times 2^{79}$ & ${*}$1.53 $\times 2^{72}$ & ${*}$1.02 $\times 2^{152}$\\
RECTANGLE-64/128 [$\clubsuit$]   & 2 & ${*}$ 75,980  & ${*}$1.26 $\times 2^{80}$ & ${*}$1.40 $\times 2^{72}$ & ${*}$1.76 $\times 2^{152}$\\
GIFT-64/128 [$\clubsuit$]  & 2 & ${*}$ 56,308  & ${*}$1.87 $\times 2^{79}$ & ${*}$1.90 $\times 2^{72}$ & ${*}$1.78 $\times 2^{152}$ \\
PRESENT-64/128 [$\clubsuit$]   & 2 & ${*}$ 69,160  & ${*}$1.15 $\times 2^{80}$ &  ${*}$1.92 $\times 2^{72}$ & ${*}$1.11 $\times 2^{153}$ \\
PUFFIN-64/128 [$\clubsuit$]    & 2 & ${* }$ 86,960  & ${*}$1.56 $\times 2^{80}$ & ${*}$1.09 $\times 2^{73}$ &  ${*}$1.70 $\times 2^{153}$ \\
GIFT-128/128 [$\clubsuit$]  & 1 & ${* }$ 95,762  & ${*}$1.59 $\times 2^{80}$ & ${*}$1.62 $\times 2^{73}$ & ${*}$1.29 $\times 2^{154}$ \\
GIFT-128/128 [$\clubsuit$]  & 2 & ${*}$191,780  & ${*}$1.59 $\times 2^{81}$ & ${*}$1.62 $\times 2^{73}$ & ${*}$1.29 $\times 2^{155}$ \\
LiCi-64/128 [$\blacktriangle$ + $\clubsuit$]    &  2 & ${*}$ 83,196  & ${*}$1.12 $\times 2^{80}$ & ${*}$1.86 $\times 2^{74}$ & ${*}$1.04 $\times 2^{155}$ \\
DEFAULT-128/128 [$\clubsuit$]  &  1 & ${*}$128,342  & 1.59 $\times 2^{81}$ & 1.75 $\times 2^{75}$ & 1.39 $\times 2^{157}$ \\
CHAM-64/128 [$\blacksquare$] &  2 & ${*}$117,504  & 1.23 $\times 2^{81}$ & 1.00 $\times 2^{76}$ & 1.23 $\times 2^{157}$ \\
HIGHT-64/128 [$\blacksquare$]  & 2 & ${*}$247,960  & 1.38 $\times 2^{82}$ & 1.90 $\times 2^{75}$ & 1.31 $\times 2^{158}$\\
CHAM-128/128 [$\blacksquare$] &  1 & ${*}$125,968  & 1.30 $\times 2^{81}$ & 1.01 $\times 2^{77}$ & 1.32 $\times 2^{158}$ \\
LEA-128/128 [$\blacksquare$]  & 1 & ${*}$210,512 & 1.19 $\times 2^{82}$ & 1.24 $\times 2^{77}$ &  1.49 $\times 2^{159}$\\
SIMON-64/128 [$\blacktriangle$ + $\blacksquare$] & 3 & ${*}$ 34,776  & 1.03 $\times 2^{79}$ & 1.03 $\times 2^{80}$ & 1.06 $\times 2^{159}$\\
SPECK-64/128 [$\blacksquare$]   & 2 & ${*}$133,424 & 1.13 $\times 2^{81}$ & 1.28 $\times 2^{78}$ &  1.45 $\times 2^{159}$\\
SIMON-128/128 [$\blacksquare$] & 2 &${*}$ 50,688  & 1.03 $\times 2^{80}$ & 1.12 $\times 2^{80}$ & 1.15 $\times 2^{160}$\\
SPECK-128/128 [$\blacksquare$] & 1 & ${*}$161,748 & 1.81 $\times 2^{81}$ & 1.54 $\times 2^{79}$ &  1.40 $\times 2^{161}$\\

\midrule
\textbf{GFSPX} (This work)  & 2 & 131,096 & 1.06 $\times 2^{81}$ & 1.46 $\times 2^{77}$ & 1.55 $\times 2^{158}$ \\
             &   & 192,308\textsuperscript{$\ddagger$} & 1.30 $\times 2^{81}$\textsuperscript{$\ddagger$} &  & 1.91 $\times 2^{158}$\textsuperscript{$\ddagger$} \\
\textbf{GFSPX} (This work)  & 3 & 196,772 & 1.53 $\times 2^{81}$ & 1.46 $\times 2^{77}$ & 1.12 $\times 2^{159}$ \\
             &   & 257,984\textsuperscript{$\ddagger$} & 1.77 $\times 2^{81}$\textsuperscript{$\ddagger$} &  & 1.30 $\times 2^{159}$\textsuperscript{$\ddagger$} \\
\bottomrule 
\end{tabular}
$\clubsuit$: SPN, \quad $\blacktriangle$: FEISTEL, \quad $\blacksquare$: ARX \\
*Estimated using the values from Table \ref{tab2} for completeness.\\
\footnotesize $\ddagger$ SWAP gates are accounted for in the total cost via a 3-CNOT decomposition.
\end{table}

The ciphertext comparison triggering the phase inversion is implemented as a two-stage reversible quantum circuit. Initially, the temporary quantum register storing the candidate encryption results is evaluated against the known reference ciphertext via $n$ parallel CNOT gates. This register reduces to the computational basis state $\lvert 0^n \rangle$ if the candidate perfectly matches the reference. Thus, evaluating $r$ pairs simultaneously requires $n\cdot r=192$ qubits at unit depth and zero T-gate cost. Second, a $t$-fold multi-controlled NOT gate is applied to the phase qubit $\lvert - \rangle$ (Fig.~\ref{Fig13}). To execute the phase inversion for the all-zero state, the control qubits are conjugated with $2 \cdot n \cdot r = 384$ NOT gates. Using the standard fault-tolerant decomposition \cite{Wiebe2016}, a $t$-fold MCX gate ($t \ge 5$) requires $32t - 84$ T-gates. For $t = 192$, this yields $6{,}060$ T-gates, adding a logarithmic-depth layer executed once per Grover iteration. Inner-parallelization requires $2r$ cipher instances per iteration for the forward and inverse operations. Accordingly, the 20-round Toffoli gate count decomposes into $2r \cdot 18{,}672 \cdot 7 \approx 7.84 \times 10^{5}$ T-gates. Thus, the target marking operation contributes less than $1\%$ to the total T-gate requirement ($6{,}060 / 7.84 \times 10^{5} \approx 0.77\%$) which is a comparable fraction to the Clifford count. Therefore, the total resource of the oracle is dominated by the cipher instances.

Since evaluating pairs $r=3$ eliminates false positives, the search space ($N=2^{128}$) contains exactly one valid key state ($M=1$). The Grover operator must be applied iteratively for $j \approx \frac{\pi}{4} \sqrt{N/M} = \frac{\pi}{4} 2^{64}$ iterations to amplify the target measurement probability to near-unity \cite{Jacques2020}. To determine the total attack complexity, the integrated oracle resource metrics are scaled by this iteration bound. Scaling the $r=2$ oracle results in an estimated T-gate count of $1.09 \times 2^{80}$, a total Clifford count of $1.06 \times 2^{81}$ ($1.30 \times 2^{81}$ including SWAP overhead), and an execution depth of $1.46 \times 2^{77}$ (Table~\ref{oracle}). For $r=3$, these estimates increase to a T-gate count of $1.64 \times 2^{80}$ and a total Clifford count of $1.53 \times 2^{81}$ ($1.77 \times 2^{81}$ including SWAP overhead), while the execution critical depth remains bound at $1.46 \times 2^{78}$.

Under the NIST Post-Quantum Cryptography (PQC) framework, the security of symmetric primitives is evaluated by comparing their quantum attack complexity to that of AES \cite{nist2016, nist2020}. This framework quantifies resilience using a total cost metric, defined as the product of the quantum circuit's total gate count and its total depth. To model realistic physical limitations of quantum hardware, NIST imposes a \texttt{MAXDEPTH} constraint that bounds the maximum allowable circuit depth over specific operational timeframes: $2^{40}$ models the maximum sequential operations a quantum computer could execute in approximately one year, $2^{64}$ corresponds to ten years, and $2^{96}$ models a millennium. Based on these constraints, security benchmarks are established, with Level 1 (AES-128 equivalent) requiring a total attack cost of at least $2^{170}$, while Level 3 and Level 5 correspond to AES-192 ($2^{233}$) and AES-256 ($2^{298}$) equivalents, respectively (Levels 2 and 4 are reserved for hash function collision resistance, such as SHA-256 and SHA-384) \cite{Grassl2016}. The total quantum cost of the Grover attack on GFSPX is estimated at $1.55 \times 2^{158}$ for $r=2$ and $1.12 \times 2^{159}$ for $r=3$, see the last column of Table \ref{oracle}. Because these costs strictly fall below the $2^{170}$ threshold within the allowable \texttt{MAXDEPTH} limits (Table \ref{tab5}), GFSPX fails to meet NIST's Level 1 post-quantum security criteria. On the other hand, GFSPX shows a comparatively high Grover attack cost while it achieves the least quantum implementation cost, as we show it in Table \ref{tab2}.

\begin{table}[h]
\centering
\caption{\texttt{MAXDEPTH analysis of Grover's algorithm}}\label{tab5}
\begin{tabular}{lccccc}
\toprule
CIPHER & $r$ & $2^{40}$ & $2^{64}$ & $2^{96}$ & approximation \\
\midrule
\textbf{GFSPX}-64/128 & 2 & 118 & 94 & 62 & $2^{158}$/\texttt{MAXDEPTH}\\
\textbf{GFSPX}-64/128 & 3 & 119 & 95 & 63 & $2^{159}$/\texttt{MAXDEPTH} \\
\bottomrule 
\end{tabular}
\end{table}

\section{Conclusion}\label{sec:conclusion}

This study presents a qubit-optimized quantum circuit realization and a quantitative Grover cryptanalysis of GFSPX, a 64-bit lightweight block cipher utilizing a 128-bit key. By translating its hybrid 4-branch generalized Feistel architecture, which integrates both ARX and SPN components into a reversible quantum framework, we systematically evaluated the cipher's post-quantum resilience against a Grover key-recovery attack under the NIST framework.

To minimize resource overhead, we used the inherent reversibility of the generalized Feistel structure for an ancilla-efficient design. The $F_1$ function utilizes a modulo-$2^{n}$ ripple-carry adder for modular addition, while $F_2$ integrates S-box synthesis with a parallelized SWAP-based permutation layer. Additionally, the dynamic key schedule is optimized in-place using a depth-efficient 113-bit rotation and parallel S-boxes. Consequently, the proposed architecture occupies $209$ qubits, achieving a baseline quantum cost of $32{,}498$  ($47{,}789$ including SWAP decomposition) at a circuit depth of $7{,}617$.

Contextualizing these results within the broader landscape of lightweight BCs (Table~\ref{tab2}) reveals several key structural insights. GFSPX exhibits the lowest cost-to-depth ratio $(4.27)$ in the comparison set, reflecting a high degree of parallel execution that distributes the gate workload across the temporal depth of the circuit. Such a configuration is unfavorable under NIST \texttt{MAXDEPTH} constraints, yet remains practical for fault-tolerant hardware restricting simultaneous physical gate operations. Furthermore, categorizing the comparison set reveals two distinct cost-to-depth clusters: SPN designs (PRESENT, GIFT, RECTANGLE, PUFFIN) typically range from $45$ to $100$, while ARX constructions (SIMON, SPECK, CHAM, HIGHT, LEA) occupy the $5$ to $25$ range. GFSPX falls below both, indicating that its hybrid nature accumulates the depth contributions of both paradigms. However, its CCNOT-to-total-gate ratio ($\approx 14\%$) mirrors ARX designs as SIMON, indicating that the modular addition primitive accounts for the nonlinearity budget without introducing Toffoli overhead to the SPN branch. Finally, the 209-qubit is offset by a higher computational density per qubit, maintaining overall architectural efficiency.

To evaluate key-recovery resistance, we constructed a Grover oracle employing an inner-parallelization strategy. Given the $128/64$ key-to-block ratio, evaluating $r=3$ plaintext-ciphertext pairs is required to eliminate spurious key candidates. The composite oracle requires $628$ qubits, where the multi-controlled phase-marking step accounts for less than $1\%$ of the total T-gate cost. Scaling by the optimal $\tfrac{\pi}{4}\,2^{64}$ iterations yields a total quantum attack cost of $1.55 \times 2^{158}$ for $r=2$ and $1.12 \times 2^{159}$ for $r=3$, with a  depth of $1.46 \times 2^{77}$. Transitioning from $r=2$ to $r=3$ increases the total cost by a factor of $1.44$ while reducing the expected number of false key candidates to approximately $2^{-64}$.

Comparing these attack costs (Table \ref{oracle}) reveals the quantitative signature of the hybrid design. ARX based ciphers such as SPECK-$64/128$ and SIMON-$64/128$ exhibit Grover costs in the range of $1.0$--$1.5 \times 2^{159}$, while SPN ciphers such as GIFT-$64/128$ and PRESENT-$64/128$ cluster around $1.1$--$1.8 \times 2^{152}$, separated by six orders of magnitude in base two. The value of $1.12 \times 2^{159}$ obtained for GFSPX places the cipher within the ARX cluster despite the presence of an SPN branch responsible for rapid diffusion. This indicates that the modular addition primitive contributes the quantum-resistance characteristic associated with ARX paradigms and primarily determines the Grover-attack cost of the hybrid construction. The gap between the obtained cost of $1.12 \times 2^{159}$ and the NIST Level~1 threshold of $2^{170}$ amounts to a factor of approximately $2^{11}$, suggesting that incremental modifications to the round function are insufficient to attain Level~1 security at a 128-bit key length within the present design space. The relative overhead with SWAP decomposition, which increases the total cost from $1.12 \times 2^{159}$ to $1.30 \times 2^{159}$ for $r=3$, amounts to $\approx 16\%$ and is comparatively lower than observed in SWAP-intensive designs, an outcome that can be attributed to the parallel decomposition of the bit-level permutation in $F_2$. Lastly, since the multi-controlled phase-marking layer accounts for less than one percent of the overall T-gate cost, the reported attack estimates provide a constrained bound, leaving minimal margin for further optimization of the oracle envelope within the inner-parallelization framework.

In conclusion, these results highlight a fundamental trade-off in lightweight cryptography design. While hybrid ARX-SPN constructions provide classical hardware efficiency and substantial quantum resilience, a 128-bit key remains structurally insufficient to achieve NIST Level~1 security under realistic quantum resource constraints. Consequently, our findings suggest two main research directions. First, extending the GFSPX framework to longer key lengths (e.g., 192 or 256 bits) or augmenting it with additional ARX layers. Second, optimizing the reversible modular-addition primitive through depth-reduced or T-count-minimized adder variants would further refine the quantum resource estimates and security margins. The methodology developed in this study provides a rigorous framework for the quantum-aware evaluation of other hybrid ARX-SPN designs in resource-constrained environments.
  
\section*{Data Availability}

All data generated or analysed during this study are included in this published article and its supplementary information files.

\bibliography{sn-bibliography}

\end{document}